\pdfoutput=1 
\documentclass[%aip,apl,reprint
superscriptaddress,
 aps,
reprint
]{revtex4-1}
\usepackage{graphicx}% Include figure files
\usepackage[utf8]{inputenc}
\usepackage[T1]{fontenc}
\usepackage{mathptmx}
\usepackage{dcolumn}
\usepackage{caption}
\usepackage{ragged2e}
\usepackage{amsmath}
\usepackage{bm}% bold math
\usepackage{amsfonts}
\usepackage{amsmath}
\usepackage{amssymb}
\usepackage{graphicx}
\usepackage{color}
\usepackage{verbatim}
\usepackage{lipsum}
\usepackage{dsfont}
\usepackage{hyperref}
\usepackage{xfrac}
\usepackage[normalem]{ulem}

\usepackage{color}
\usepackage{amsmath}%[fleqn]
\usepackage{wasysym}
\usepackage{float}
\usepackage{array,url,kantlipsum}
\usepackage{verbatim}
\usepackage{xcolor}
\usepackage{CJK}
\usepackage{textcomp}

\draft % marks overfull lines with a black rule on the right
\begin{document}
\preprint{AIP/123-QED}
%\preprint{}
\captionsetup{justification=raggedright,singlelinecheck=false}
\title{Battery characterization via eddy-current imaging with nitrogen-vacancy centers in diamond} %Title of paper



\author{Xue Zhang}
\affiliation{Johannes Gutenberg-Universit{\"a}t Mainz, 55128 Mainz, Germany}
 \affiliation{Helmholtz-Institut, GSI Helmholtzzentrum f{\"u}r Schwerionenforschung, 55128 Mainz, Germany}
%\thanks{}
\author{Georgios Chatzidrosos}
\affiliation{Johannes Gutenberg-Universit{\"a}t Mainz, 55128 Mainz, Germany}
 \affiliation{Helmholtz-Institut, GSI Helmholtzzentrum f{\"u}r Schwerionenforschung, 55128 Mainz, Germany}
\author{Yinan Hu}
\affiliation{Johannes Gutenberg-Universit{\"a}t Mainz, 55128 Mainz, Germany}
 \affiliation{Helmholtz-Institut, GSI Helmholtzzentrum f{\"u}r Schwerionenforschung, 55128 Mainz, Germany}
 \author{Huijie Zheng}
 \email[Corresponding author: ]{zheng@uni-mainz.de}
\affiliation{Johannes Gutenberg-Universit{\"a}t Mainz, 55128 Mainz, Germany}
 \affiliation{Helmholtz-Institut, GSI Helmholtzzentrum f{\"u}r Schwerionenforschung, 55128 Mainz, Germany}
 \author{Arne Wickenbrock}
 \email[Corresponding author: ]{wickenbr@uni-mainz.de}
\affiliation{Johannes Gutenberg-Universit{\"a}t Mainz, 55128 Mainz, Germany}
 \affiliation{Helmholtz-Institut, GSI Helmholtzzentrum f{\"u}r Schwerionenforschung, 55128 Mainz, Germany}
 \author{Alexej Jerschow}
\affiliation{Department of Chemistry, New York University, New York, NY 10003, USA} 
 \author{Dmitry Budker}
\affiliation{Johannes Gutenberg-Universit{\"a}t Mainz, 55128 Mainz, Germany}
 \affiliation{Helmholtz-Institut, GSI Helmholtzzentrum f{\"u}r Schwerionenforschung, 55128 Mainz, Germany}
\affiliation{Department of Physics, University of California, Berkeley, California 94720, USA}

\date{\today}

\begin{abstract}
Sensitive and accurate diagnostic technologies with magnetic sensors are of great importance for identifying and localizing defects of rechargeable solid batteries in a noninvasive detection. We demonstrate a microwave-free AC magnetometry method with negatively charged NV centers in diamond based on a cross-relaxation feature between NV centers and individual substitutional nitrogen (P1) centers occurring at 51.2\,mT. We apply the technique to non-destructive solid-state battery imaging. 
%magnetic sensitivity is achieved $40nT/\sqrt{Hz}$ with $100kHz$ bandwidth. 
By detecting the eddy-current-induced magnetic field of the battery, we distinguish a defect on the external electrode and identify structural anomalies within the battery body. The achieved spatial resolution is %$360\,(2)\,\mu\rm m$ at a 100\,(0.5)\,$\mu$m 
$360\,\mu\rm m$. The maximum magnetic field and phase shift generated by the battery at modulation frequency of 5\,kHz are estimated as 0.04\,mT and 0.03\,rad respectively.
%\DB{I do not understand what you mean here. Also, please leave a small space between numbers and units and do not make units italic.}  
\end{abstract}


\maketitle %\maketitle must follow title, authors, abstract and \pacs


\section{Introduction}
Sensitive and accurate diagnostic technologies with magnetic sensors are of great importance for identifying and localizing defects of rechargeable solid batteries in a noninvasive detection. Negatively charged nitrogen-vacancy (NV) centers have been extensively exploited as precise nanoscale probes in applications such as measurement of magnetic fields\, \cite{Wickenbrock2016_MF}, temperature \,\cite{Kucsko2013_tem_depen_NV}, strain\,\cite{Ovartchaiyapong2014_strain_effect_NV}, rotation\, \cite{Ajoy2012_rotation_NV, Ledbetter2012_gyroscope_NV}, electric fields\, \cite{Dolde2011_Electricfield_NV, Block2020_Electricfield_NV} and so forth. 
%Due  to the combination of high sensitivity and spatial resolution, 
Recently, NV centers have also been utilized for eddy-current imaging. Eddy-current detection was demonstrated with vapor-cell magnetometers\,\cite{Wickenbrock2014_vaporcell,Marmugi2016_vaporcell,Wickenbrock2016_vaporcelleddy},  and later also with NV diamond\, \cite{Chatzifrosos2019_eddycurrent_NV}. Compared with other sensors, diamond-based devices can be used over a wide temperature range, can have nanoscale spatial resolution, high sensitivity and wide bandwidth. In ref.\,\cite{Chatzifrosos2019_eddycurrent_NV}, a microwave-free eddy-current imaging device based on NV centers in diamond was demonstrated making use of a NV-NV cross-relaxation feature between 0 and 20\,mT. In this paper, we report imaging based on a much narrower cross-relaxation feature between the NV centers and P1 centers and apply it to perform non-destructive evaluation experiments on small solid-state batteries. 

The imaged sample is an all-ceramic multilayer solid-state battery produced by TDK Corporation. It incorporates inner electrodes, electrolyte and external electrodes. When the battery is placed in an oscillating magnetic field (primary field), eddy currents flow in the electrodes and the electrolyte, which, in turn in our case induce an alternating magnetic field (secondary field). 

We demonstrate all-optical AC magnetometry to detect the secondary field as a function of position. The secondary field is anti-parallel to the primary field and exhibits a phase delay. These quantities relate to a number of characteristics in the imaged sample, such as: shape, dimensions, conductivity and susceptibility. As a result, one can use NV centers as probes to distinguish textures and identify structural anomalies in- and outside the battery.

\section{Experimental apparatus}
\begin{figure*}
\includegraphics[scale=1.75]{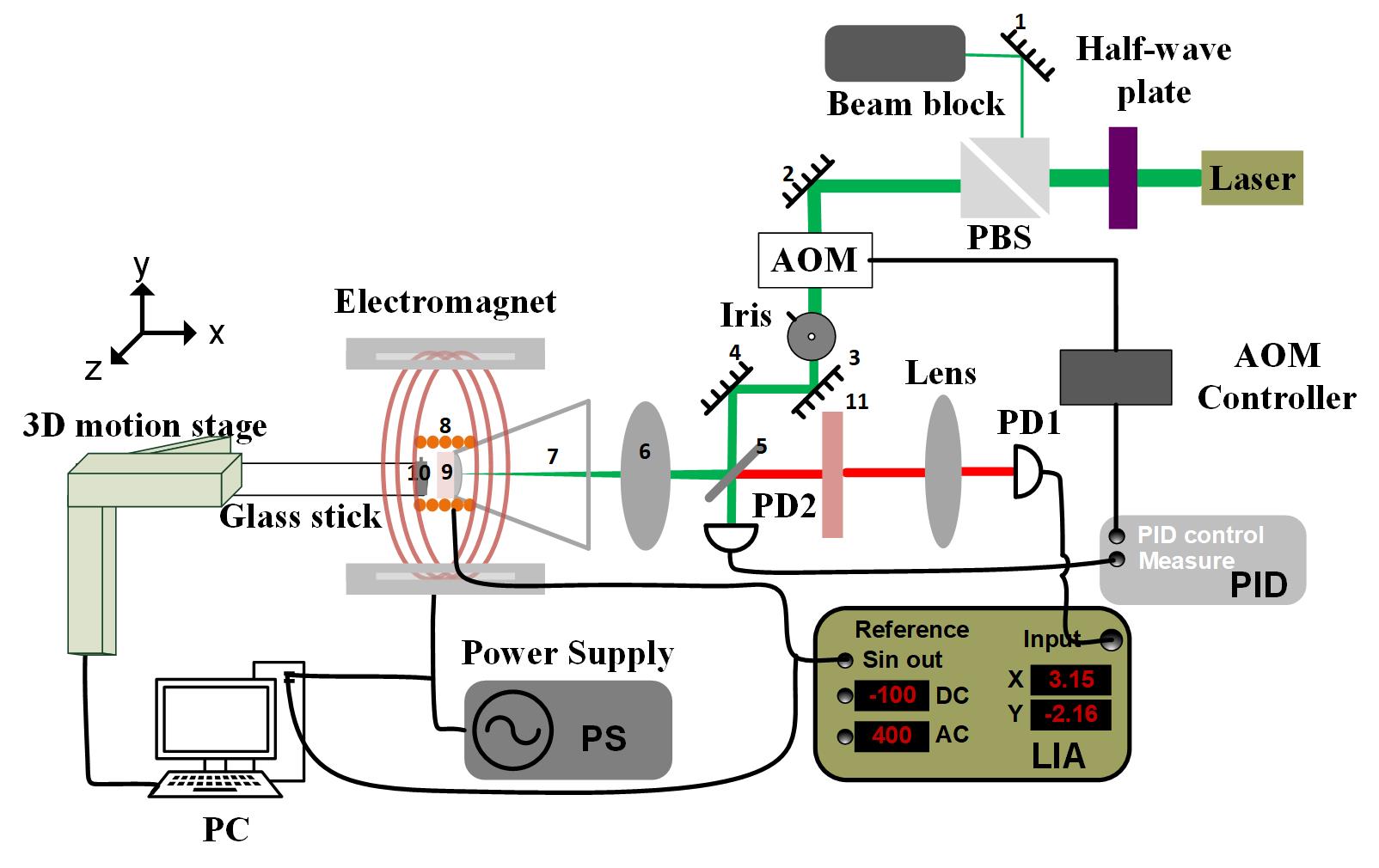}
\caption{Experimental setup. 1-4, mirror 5, dichroic mirror 6, lens 7, Parabolic concentrator 8, RF coil 9, diamond 10, battery 11, Notch filter. PBS: polarizing beamsplitter; AOM: acousto-optic modulator; PD1,2: photodiodes; PC: personal computer; PS: power supply.}
\label{Fig:experimental_setup}
\end{figure*}
A schematic of the experimental setup is shown in Fig.\,\ref{Fig:experimental_setup}. Linearly polarized light from a continuous-wave green laser at 532\,nm is continuously pumping the NV centers in a diamond sample. The laser power can be adjusted with a polarizer and a half-wave-plate. NV centers in the excited state decay back to the ground state both by radiative and nonradiative transitions, with red photoluminescence (PL) emitted in the former case. The fact that NV centers in the $m_s=\pm1 $ magnetic sublevels have significantly higher probability to undergo an intersystem crossing (ISC)\, \cite{Acosta2010_ISC}, i.e a nonradiative transition from the triplet to the singlet excited state, results in the majority of NV centers being optically pumped into the $m_s=0$ spin projection. In order to minimize the noise caused by intensity fluctuations in the laser beam, we use a light-power stabilization loop incorporating a photodiode (PD2), a proportional-integral-differential (PID) controller, and  an acousto-optic modulator (AOM). 

A dichroic mirror and a notch filter are used to filter out red PL from the green light. The diamond used in this setup is a type-Ib, (111)-cut, high-pressure-high-temperature-grown sample with dimension $3.0\times 3.0 \times 0.4 \, \rm mm^3$. The initial nitrogen concentration was specified less than 110\,ppm. Then nitrogen-rich sample was electron-irradiated and annealed\,\cite{Chatzifrosos2019_eddycurrent_NV}. The diamond is glued on the top plane of a parabolic concentrator used to collect fluorescence. A radio-frequency (RF) coil is made from copper wire of 0.1\,mm diameter. It has five turns and is wound around the diamond to provide a modulation field to both the diamond and battery. The modulation field is provided by the Lock-in amplifier (LIA) used for the AC magnetometric scheme. The frequency of the field (maximum 4 MHz) is also used as the LIA reference for detection.

A 3D translation stage is used to raster scan the battery in front of the diamond. To reduce the effect of magnetic noise produced by the 3D translation stage. The battery is glued on top of a $24\,\rm{cm}$ rod made from polyvinyl chloride (PVC) onto a motorized three-dimension-translation-stage controlled by a computer. A variable background field is provided by a custom-made electromagnet (EM). This field is required for the microwave-free magnetic field detection method. The EM consists of approximately 200 turns wound with a rectangular cross-section ($1.4 \times 0.8 \,\rm{mm^2}$) wire around a 5 cm-diameter bore. The coil is wound on a water-cooled copper mount, and produces a background field of $2.9\,\rm{mT}$ per ampere supplied. Diamond, battery and RF coil are all placed inside the bore of the EM. A LIA detects the amplitude ($R$) and phase ($\theta$) of PL modulation. The LIA is connected to the computer, and $R$ and $\theta$ are recorded along with position of the battery (the 3D translation stage).
\begin{figure*}
    \centerline{
    \includegraphics{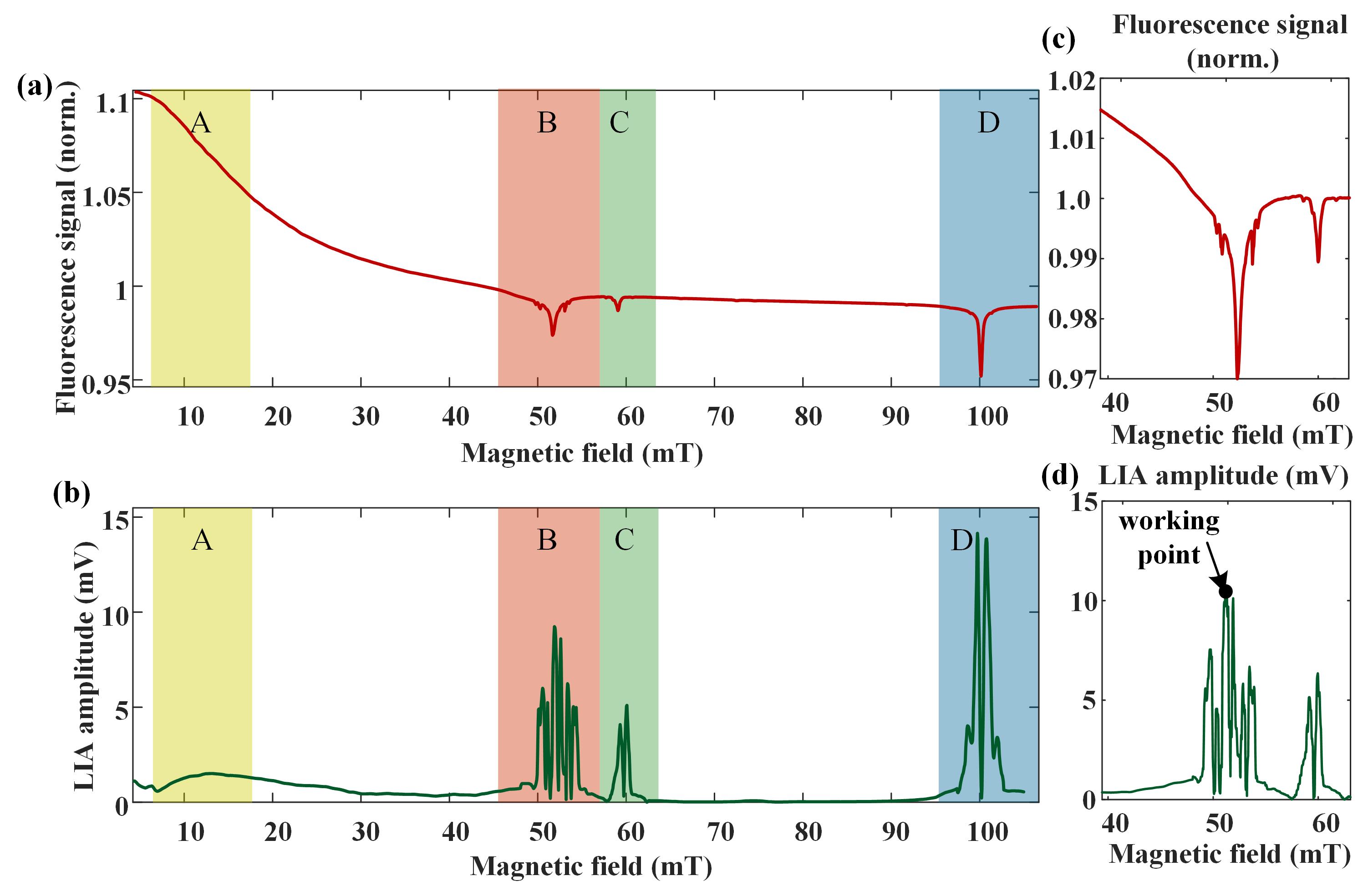}
    }
    %\renewcommand{\raggedright}{\leftskip=0pt \rightskip=0pt plus 0cm}
   % \captionsetup{width=\textwidth}
    \caption{(a). PL of NV centers as a function of the background magnetic field. The four areas A, B, C, D that can be exploited for magnetometry indicate features due to spin mixing, cross relaxation between NV centers and P1 centers, cross relaxation with NV centers that are not aligned along the magnetic field, and ground state level anticrossing, respectively. (b) Corresponding amplitude of LIA referenced at 100\,Hz for those features. (c) Zoomed-in B area of (a), PL recorded with field-modulation depth of 0.48\,mT at 100\,Hz. (d) Zoomed-in B area of (b), indicating the working point at 52.5\,mT.
    \label{Fig:PL}}
\end{figure*}

\section{AC magnetometry}
The PL of a diamond sample displays various cross relaxation features, extensively discussed in the literature\,\cite{Zheng2017_gslac, Hall2016_features_NV, Anishchik2017_features_NV}, as a function of the bias magnetic field, as shown in Fig.\,\ref{Fig:PL}. In this letter, we exploit a cross-relaxation feature between NV and P1 centers occurring at 51.2\,mT [area B in Fig.\,\ref{Fig:PL}\,(a)]. This feature is used in the battery measurement because of its improved sensitivity compared to the broad NV-NV cross-relaxation feature near zero field (area A), and of its reduced dependence on magnetic field compared to cross relaxation with NV centers not aligned along the magnetic field at 60\,mT (area C) and to the ground state level-anticrossing (GSLAC) feature at 102.4\,mT (area D). Figure\,\ref{Fig:PL}\,(b) shows the corresponding signal amplitude $R$ detected with the LIA, demodulated at the referenced frequency. To maximize the response to alternating signals, $R$ has to be maximized. The NV features seen in Fig.\,\ref{Fig:PL} are expected to display a temperature dependence. For example, the NV-P1 cross-relaxation features are expected to shift by $-1.34\, \mu$T/K at room temperature\,\cite{Acosta2010_temdepen_NV}, which can lead to drifts and be interpreted as magnetic field noise. 
As noted in the inset (d), we choose to use the features at 52.5\,mT
%which are more robust against temperature change
to detect the AC magnetic signals. The noise sensitivity of the sensor for the parameters used in this experiment leads to an estimated sensitivity of $40\,\rm {nT/\sqrt{Hz}}$ with a bandwidth of 100\,kHz. It should be pointed out that the bandwidth of this technique can be expanded up to MHz, depending on pump-power intensity and NV-axis alignment with respect to the external magnetic field\,\cite{Chatzifrosos2019_eddycurrent_NV}. 

\begin{figure*}
    \centerline{
    \includegraphics[scale=0.35]{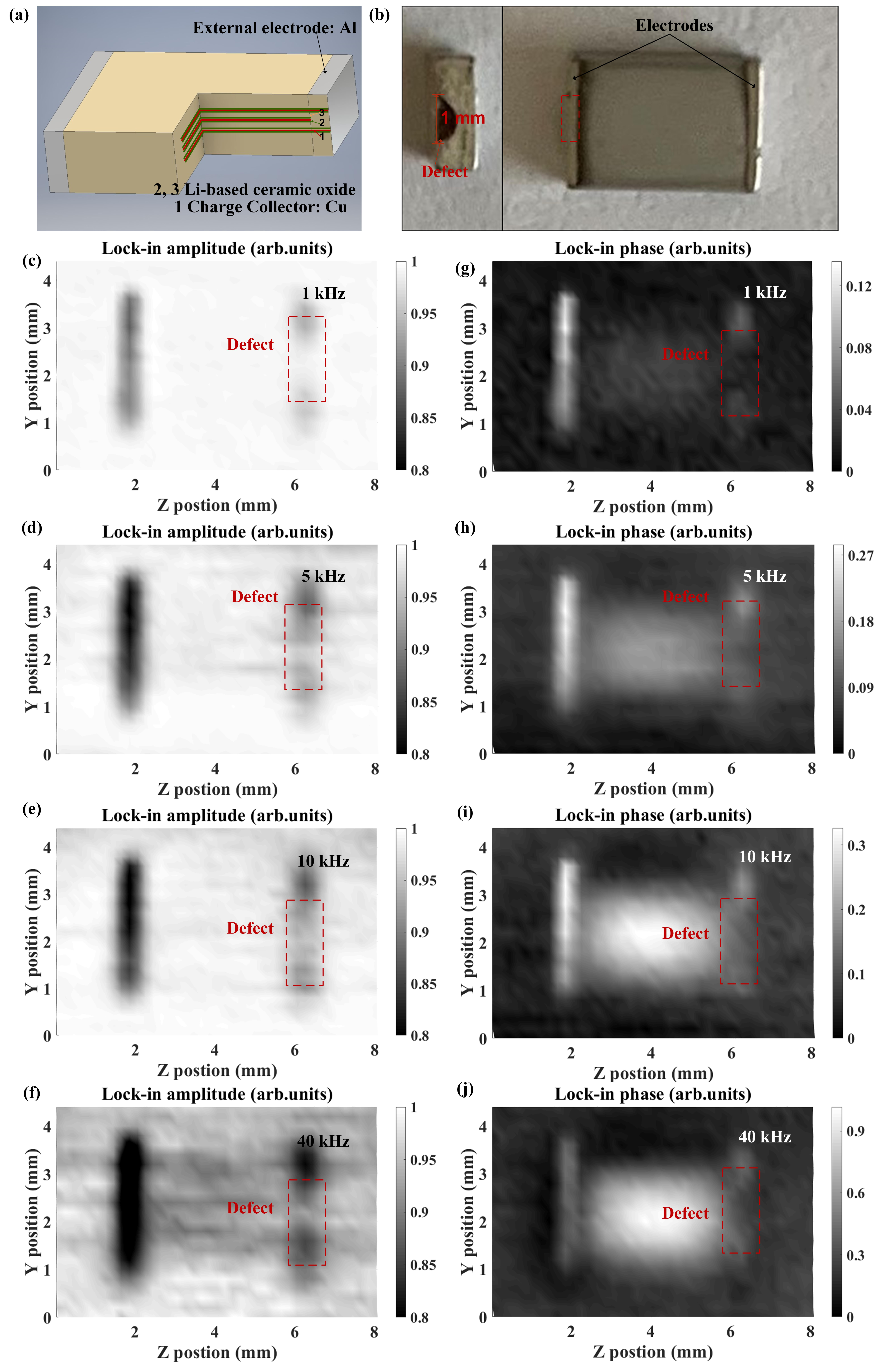}
    }
    % \captionsetup{width=\textwidth} 
    % \renewcommand{\raggedright}{\leftskip=0pt \rightskip=0pt plus 0cm}
    \caption{Continued on the following page.
    % \newline{(a) Battery schematic. The ceramic solid battery consists of external aluminium electrodes, Li-based ceramic oxide inner electrodes (noted as 2), Li-based ceramic oxide electrolyte (noted as 3), and copper charge collector (indicated as 1). (b) shows a photograph of the battery with dimension $4 \times 3 \times 1 \,\textrm{mm}^3 $. Distinctly visible is the defect on the external electrode. (c)-(f) show maps of the $\theta$ of LIA for modulation frequencies of 500\,Hz, 5\,kHz, 20\,kHz and 40\,kHz, respectively. (g)-(j) show maps of the $R$ of LIA for modulation frequencies of 500\,Hz, 5\,kHz, 20\,kHz and 40\,kHz.}
    }
    \label{Fig:map}
\end{figure*}

\begin{figure*}
    \ContinuedFloat
   % \captionsetup{width=\textwidth} 
    %\renewcommand{\raggedright}{\leftskip=0pt \rightskip=0pt plus 0cm}
    \caption{(a) Battery schematic. The ceramic solid battery consists of external aluminium electrodes, Li-based ceramic oxide inner electrodes (noted as 2), Li-based ceramic oxide electrolyte (noted as 3), and copper charge collector (indicated as 1). 
    Image from Ref\,\cite{TDKweb}. 
    (b) A photograph of the battery with dimension $4 \times 3 \times 1 \,\textrm{mm}^3 $. Distinctly visible is the defect on the external electrode. (c)-(f) Maps of the $R$ of LIA for modulation frequencies of 1\,kHz, 5\,kHz, 10\,kHz and 40\,kHz. (g)-(j) Maps of the $\theta$ of LIA for modulation frequencies of 1\,kHz, 5\,kHz, 10\,kHz and 40\,kHz, respectively.}
\end{figure*}

\section{battery measurement}
The dimensions of the solid state battery measured in this work is $4.0 \times 3.0 \times 1.0 \,\rm{mm^3}$, and are shown in Fig.\,\ref{Fig:map}\,(a)\cite{hu2020rapid}. It comprises Li-based-ceramic-oxide inner electrodes (noted as 2), Li-based-ceramic-oxide electrolyte (denoted as 3), copper charge collector (indicated as 1), and external electrodes made of aluminum. Figure\,\ref{Fig:map}\,(b) shows a photograph of the solid battery, which has an artificial 1 mm-length defect at the left external electrode. The battery sample is placed at distance of 0.1 mm to the diamond at the perpendicular direction ($x$-direction), and scanned over the transverse plane ($y$-$z$ plane). Figure\,\ref{Fig:map}\,(g)-(j) show the spatially resolved phase signals of the LIA at the corresponding modulation frequencies of 1\,kHz, 5\,kHz, 10\,kHz, 40\,kHz, respectively. The external electrodes including the defect are distinctly visible under the modulation frequency of $1\,\rm{kHz}$ as shown in Fig.\,\ref{Fig:map}\,(g). As the modulation frequency is increasing, the phase $\theta$ shows an image of the battery structure between the electrodes. There the induced field from different portions of the battery can be differentiated, as shown in Fig.\,\ref{Fig:map}\,(h)-(j). Figure\,\ref{Fig:map}\,(c)-(f) show the spatially resolved amplitude signals of LIA at the corresponding modulation frequencies of 1\,kHz, 5\,kHz, 10\,kHz, 40\,kHz.

\begin{figure*}
    \centerline{
    \includegraphics[scale=0.35]{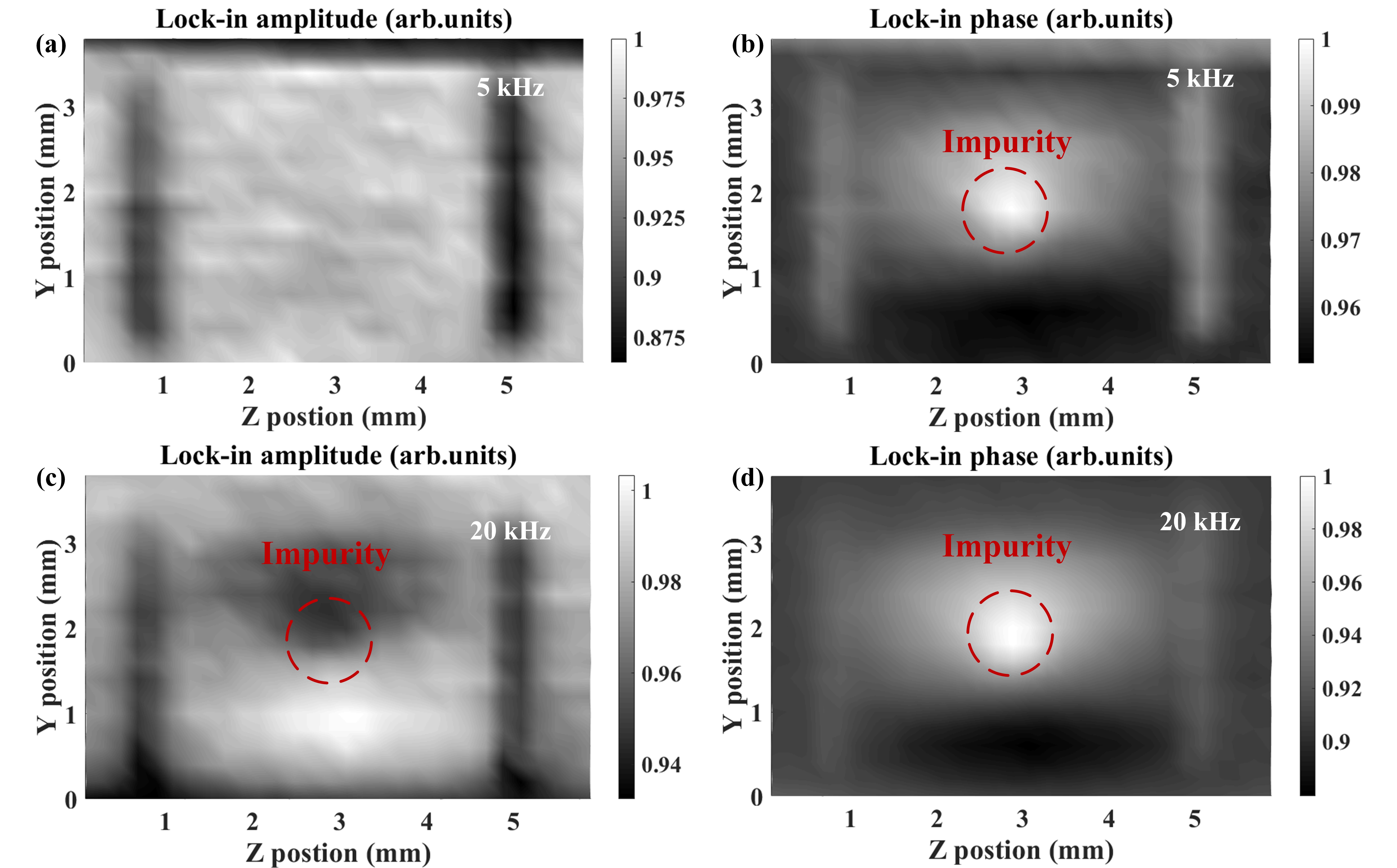}
    }
   \captionsetup{width=\textwidth} 
    \renewcommand{\raggedright}{\leftskip=0pt \rightskip=0pt plus 0cm}
    \caption{Imaging pictures of solid batteries with the impurity. (a) and (b), (c) and (d) are amplitudes and phases of LIA for the battery with an impurity at frequencies of 5\,kHz and 20\,kHz respectively. (e) and (g), (f) and (h) are amplitudes and phases of LIA for the battery with a crevice in the middle at frequencies of 5\,kHz and 20\,kHz respectively.
    }
    \label{Fig:impurity}
\end{figure*}

\begin{figure*}
    \centerline{
    \includegraphics[scale=0.35]{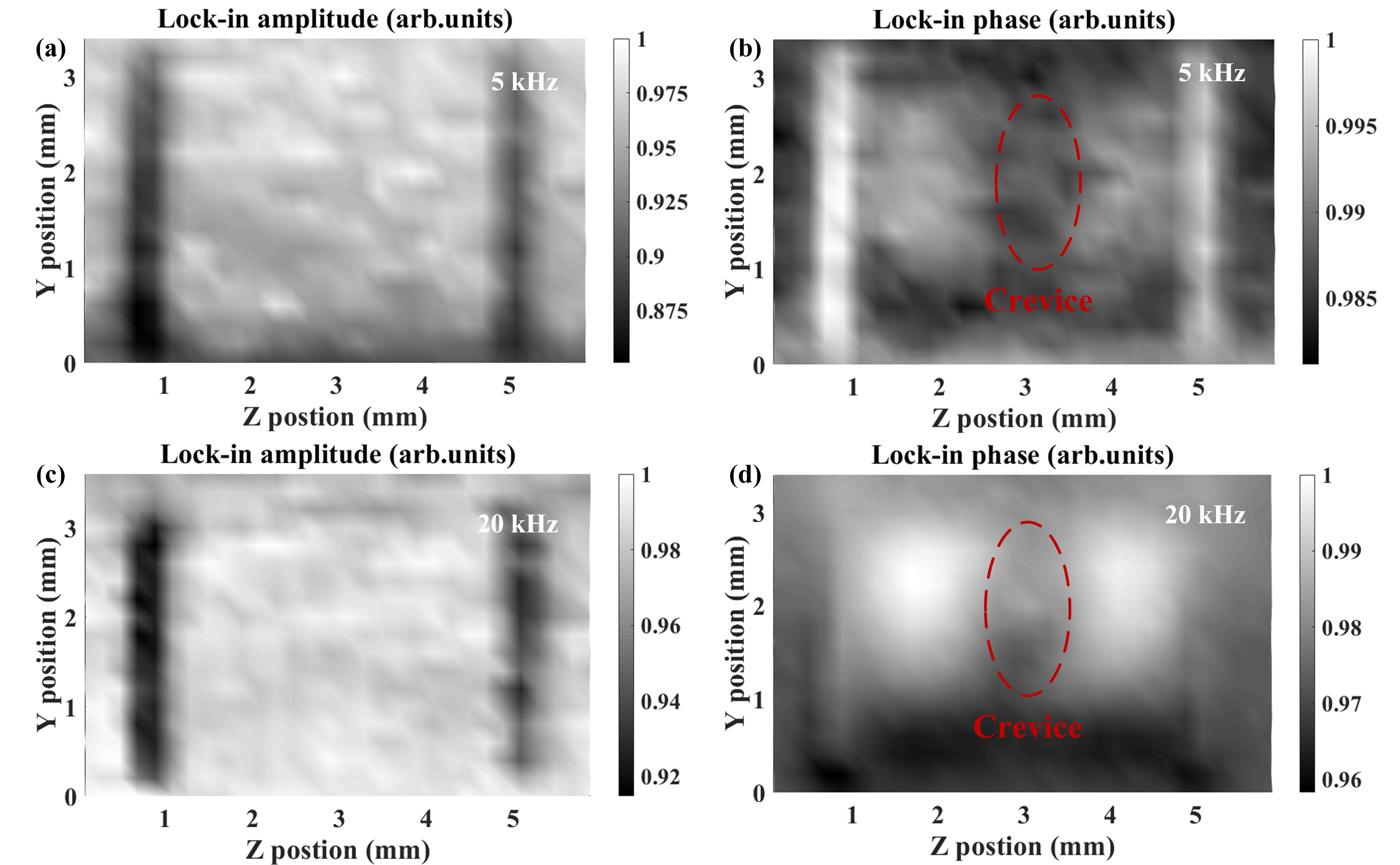}
    }
   % \captionsetup{width=\textwidth} 
   % \renewcommand{\raggedright}{\leftskip=0pt \rightskip=0pt plus 0cm}
    \caption{Imaging pictures of solid batteries with the crevice. (a) and (b), (c) and (d) are amplitudes and phases of LIA for the battery with a crevice at frequencies of 5\,kHz and 20\,kHz respectively.
    }
    \label{Fig:crevice}
\end{figure*}

In order to investigate the ability of diamond-based magnetometer to identify structural anomalies inside batteries, we introduced two kinds of artificial anomalies. First, an iron-containing brass cylinder with a height of 1\,mm and diameter of 1\,mm is placed inside the battery, serving as an impurity. The results are shown in Fig.\,\ref{Fig:impurity}\,(a)-(d), displaying the spatially resolved amplitude and phase signals of the LIA for the solid battery at frequencies of 5\,kHz and 20\,kHz respectively. This impurity is visible at both frequencies in the phase signal, indicated with red dashed circles in Fig\,\ref{Fig:impurity}. The impurity image is absent in the amplitude signal at frequencies lower than 5\,kHz, this is because the amplitude variations caused by the impurity, are smaller than the background shifts caused by temperature changes, for the amplitude signal. The other anomaly is made by cutting the battery in half and then attaching the halves back together, leaving no defect visible to the naked eye. The results are shown in Fig\,\ref{Fig:crevice}, where (a)-(d) display spatially resolved amplitude and phase signals of LIA for the solid battery at frequencies of 5\,kHz and 20\,kHz respectively. The crevice is distinctly visible in the middle of the battery in the phase signals at both frequencies, indicated by the yellow ellipses in (b) and (d). In the amplitude signals the crevice is absent again. 

%For further determination of the magnetic field generated by the solid state battery, 
The primary field produced at the center of the coil where the battery is located is given by:  
%the induced field provided by RF coil can be calculated by:
%\begin{equation}
%B_{\rm{primary}}=\frac{\mu_0*N*I*r_{\rm{coil}}^2}{2(r_{\rm{coil}}^2+d^2)^\frac{3}{2}}\label{(1)},
%\end{equation}
\begin{equation}
B_{\rm{primary}}=\frac{\mu_0\times N\times I}{2r_{\rm{coil}}},
\label{(1)}
\end{equation}
where $\mu_0=4\pi\times10^{-7} \,\rm{T\times m/A}$ is the permeability of free space, $N=4$ is the number of turns of the RF coil, $I=0.8\,\rm A$ is the amplitude of the alternating current through the coil, and $r_{\rm{coil}}=4.2\,$mm is the radius of the coil.
%, and $d=0.1\,\rm{mm}$ represents the distance between the battery and the  diamond. 
To perform the LIA detection scheme the modulated field amplitude is selected to be 0.48\,mT, confirmed experimentally by the GSLAC PL scan and theoretically by eq.\ref{(1)}. 

\begin{figure*}
    \centerline{
    \includegraphics[scale=0.35]{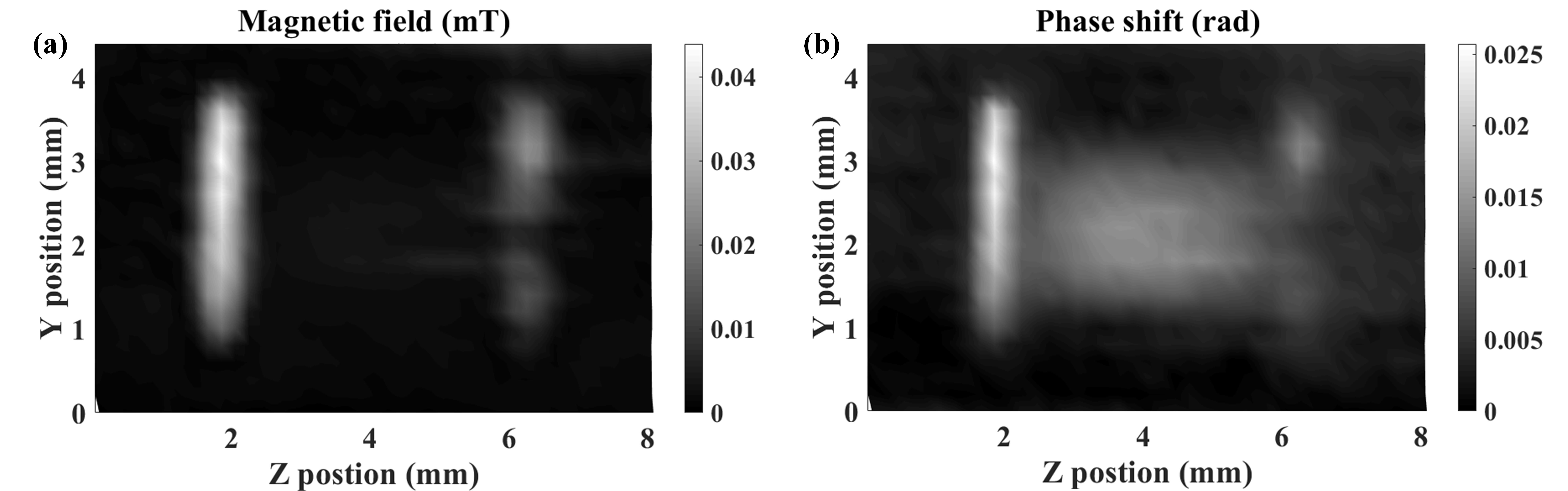}
    }

        \caption{Magnetic field and phase shifts generated by the battery at frequency of 5\,kHz. (a) Calibrated magnetic field generated by the battery. (b) Phase shifts generated by the battery. }
    \label{Fig: MF}
\end{figure*}

For AC measurements, the magnetic susceptibility $\chi_m$ yields two quantities\,\cite{Gomory1997_ACsusceptibility, Mulder1981_ACsusceptibility}: the magnitude and phase of the susceptibility. Alternatively, we can express it in the form of complex susceptibility with the real component $\chi_r$ and imaginary component $\chi_i$,
\begin{equation}
\chi_m=\chi_r+i\chi_i
\end{equation} 
where $\chi_r$ is the slope of DC magnetization curve $M(H)$ of the battery, and $\chi_i$ indicates dissipative processes in the battery. %Based on that, the magnetic field generated by the battery can be expressed by
%\begin{equation}
%\begin{split}
%&B_{\rm{battery}}=\chi_rB_{\rm{primary}}+\\
%&i\int_{0}^{a}\int_{0}^{b}\int_{0}^{h}\mu_0 (1+\chi_i)\sigma fB_{\rm{primary}}xy(\frac{x}{(x/2)^2+(z-d-h)^2}-\frac{y}{(y/2)^2+(z-d-h)^2})dxdydz
%\end{split}
%\end{equation}
%where $a,b,h$ are length, width and depth of the solid battery respectively, $\sigma$ is the mutual conductivity of the solid battery, $f$ represents the frequency of inducing B-field. Precisely, the integral depth $h$ is relevant to the skin depth $\delta$ of inducing field,
The secondary field produced by the battery also depends on the skin depth in the conducting material: 
\begin{equation}
\delta=\sqrt{\frac{2}{\sigma\omega\mu}} ,
\end{equation}

where $\omega=2\pi f$ denotes the angular frequency, $\sigma$ is conductivity of the material and $\mu$ is mutual permeability of the material. In order to estimate the AC magnetic field generated by the battery, we adopt the calibration formula in ref\,\cite{Wickenbrock2014_vaporcell},
\begin{equation}\label{eq:12}
\begin{split}
& \delta\phi=\arctan{(\frac{R_0\cos{(\phi_0)}-R_m\cos{(\phi_m)}}{R_m\sin{(\phi_m)}})},\\
& \delta R=\sqrt{R_m^2\sin{(\phi_m)}^2+(R_0\cos(\phi_0)-R_m\cos{(\phi_m)})^2}, \\
\end{split}
\end{equation}

% \begin{gather}
%     \delta\phi=\arctan{(\frac{R_0\cos{(\phi_0)}-R_m\cos{(\phi_m)}}{R_m\sin{(\phi_m)}})},\\
% \delta R=\sqrt{R_m^2\sin{(\phi_m)}^2+(R_0\cos(\phi_0)-R_m\cos{(\phi_m)})^2},
% \end{gather}
where $\delta\phi$ is phase shifts generated by the battery, $\delta R$ is amplitude contrast generated by the battery, $R_0$ and $\phi_0$ are the background-corresponding amplitude and phase of the LIA, $R_m$ and $\phi_m$ are the measured lock-in amplitude and the phase. The calibrated results are shown in Fig\,\ref{Fig: MF}. We calibrated the magnetic field and phase shifts generated by the battery in Fig\,\ref{Fig:map} at modulation frequency of 5\,kHz. The external electrode without the defect induces the maximum magnetic field of 0.04\,mT and phase shift of 0.03\,rad.   

\section{Spatial resolution}
\begin{figure}
    \centerline{
    \includegraphics[scale=0.45]{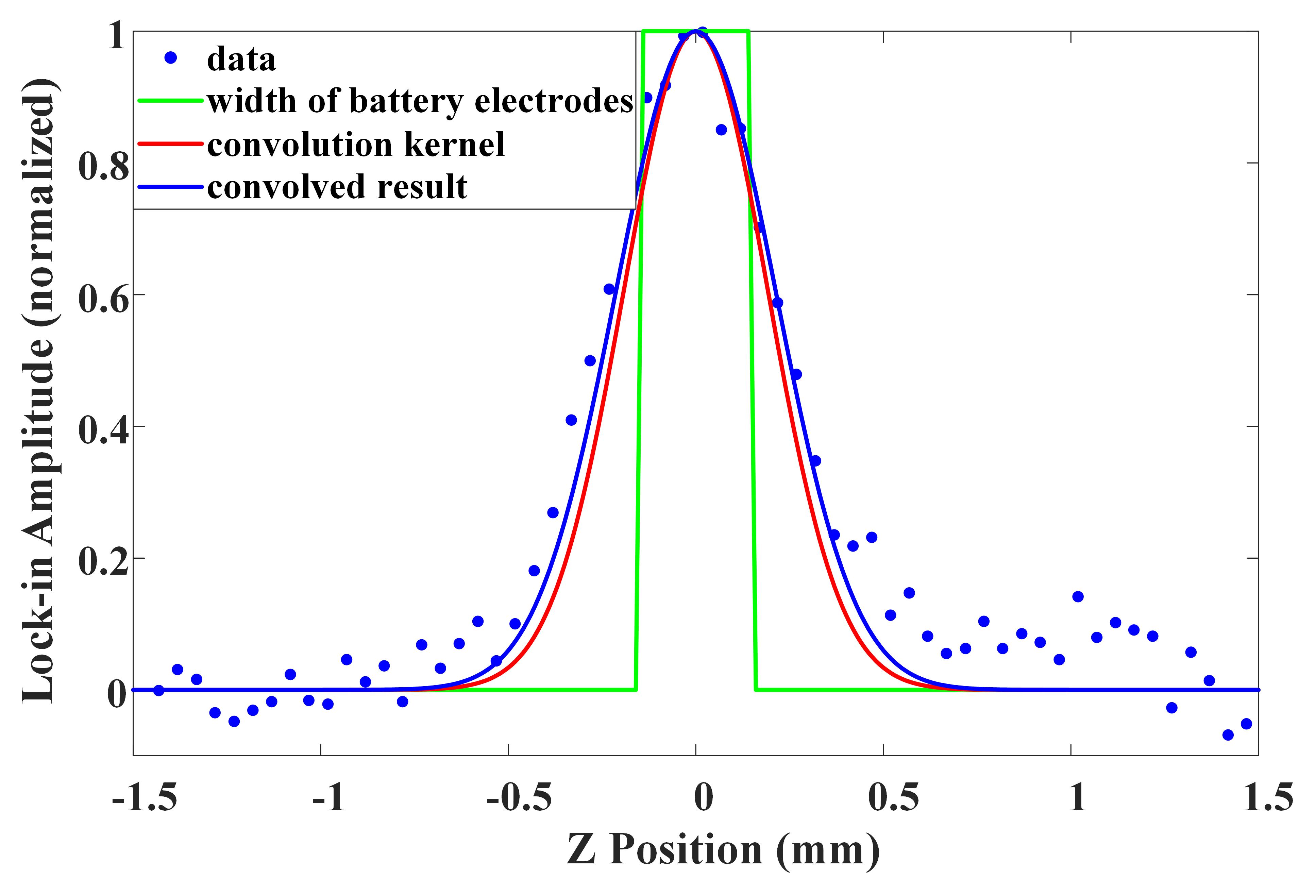}
    }
    \renewcommand{\raggedright}{\leftskip=0pt \rightskip=0pt plus 0cm}
        \caption{Determination of spatial resolution. The blues dots signify the average cross section of the 0.3 mm-width electrode imaged by the sensor, the green trace denotes the width of electrode of battery as a square function, and the red curve represents a Gaussian function, as a convolution kernel of the green trace to fit the experimental data, noted as blue trace.}
    \label{Fig: spatialresolution}
\end{figure}
One of the advantages of diamond-based sensors is the spatial resolution. For determining the spatial resolution of this battery-measurement system, we use data from one of the aluminium-made electrodes of the solid battery. As shown in Fig.\,\ref{Fig: spatialresolution}, the average cross section of 0.3 mm-width electrode is expressed by a square function, noted as the green plot, a Gaussian-function, noted with red is employed as a kernel to convolve the square function to fit the measured data, denoted as blue dots. All traces have been normalized for better comparison. By adjusting the full width at half-maximum of Gaussian function, we can estimate the spatial resolution of the system. We represent that results with blue. The spatial resolution of the system can be estimated form the full width at half maximum (FWHM) of the kernel and is estimated as $360 \pm 2\, \mu \rm m$. It is mostly restrained by the distance between the battery and diamond. Closer proximity could increase measurement contrast, allowing more legible image of small samples.

\section{Conclusion and Outlook}
We demonstrate microwave-free eddy-current imaging with NV centers in diamond exploiting the cross-relaxation feature between the NV centers and P1 centers occurring at 51.2 mT. The magnetic sensitivity is $40\,\rm{nT/\sqrt{Hz}}$ with 100 kHz bandwidth, but as previously demonstrated\cite{Chatzifrosos2019_eddycurrent_NV} it can be extended to a few MHz. The diamond based sensor is used to noninvasively image a mm-sized solid battery. The spatially resolved amplitude and phase lock-in signals can be used to determine the dimension and shape of the battery features, including a 1\,mm-wide artificial defect and 0.3\,mm-wide electrodes, from which the spatial resolution is calibrated as $360(2)\,\mu\rm m$ limited by the distance between the sensor and the battery. Furthermore, diamond-based probe also identifies an anomaly and a crevice inside the battery, demonstrating the ability to detect internal anomalies which are not visible to the naked eye. The maximum magnetic field and phase shift generated by the battery at modulation frequency of 5\,kHz are estimated as 0.04\,mT and 0.03\,rad, respectively. 

This paper mainly focuses on testing solid batteries. For future studies, the spatial resolution could be further improved by either using a thinner diamond sample or a diamond with a shallow implanted NV layer to shorten the sensor-sample distance. By tuning the bandwidth, it would be possible to identify structures, at different depth inside the battery including electrodes,  electrolytes\,\cite{hu2020sensitive} and even material components, which could be useful for battery assessment and development.

\section{Acknowledgements}

The work was funded in part by a grant by the U.S. National Science Foundation under award CBET 1804723 and the German Federal Ministry of Education and Research (BMBF) within the Quantumtechnologien program (FKZ13N14439 and FKZ 13N15064), and by EU FET-OPEN Flagship Project ASTERIQS (action 820394).

% If in two-column mode, this environment will change to single-column format so that long equations can be displayed. 
%\begin{widetext}
%$$\mbox{put long equation here}$$
%\end{widetext}

% Figures should be put into the text as floats. 
% Use the graphics or graphicx packages (distributed with LaTeX2e).
% Put the label that you will use with \ref{} command in the braces of the \label{} command.
%
% \begin{figure}
% \end{figure}

% \begin{table}
% \end{tabular}
% \end{table}

%\begin{acknowledgments}
% Put your acknowledgments here.
%\end{acknowledgments}

\bibliography{aipsamp}

\providecommand{\noopsort}[1]{}\providecommand{\singleletter}[1]{#1}%
\begin{thebibliography}{21}%
\makeatletter
\providecommand \@ifxundefined [1]{%
 \@ifx{#1\undefined}
}%
\providecommand \@ifnum [1]{%
 \ifnum #1\expandafter \@firstoftwo
 \else \expandafter \@secondoftwo
 \fi
}%
\providecommand \@ifx [1]{%
 \ifx #1\expandafter \@firstoftwo
 \else \expandafter \@secondoftwo
 \fi
}%
\providecommand \natexlab [1]{#1}%
\providecommand \enquote  [1]{``#1''}%
\providecommand \bibnamefont  [1]{#1}%
\providecommand \bibfnamefont [1]{#1}%
\providecommand \citenamefont [1]{#1}%
\providecommand \href@noop [0]{\@secondoftwo}%
\providecommand \href [0]{\begingroup \@sanitize@url \@href}%
\providecommand \@href[1]{\@@startlink{#1}\@@href}%
\providecommand \@@href[1]{\endgroup#1\@@endlink}%
\providecommand \@sanitize@url [0]{\catcode `\\12\catcode `\$12\catcode
  `\&12\catcode `\#12\catcode `\^12\catcode `\_12\catcode `\%12\relax}%
\providecommand \@@startlink[1]{}%
\providecommand \@@endlink[0]{}%
\providecommand \url  [0]{\begingroup\@sanitize@url \@url }%
\providecommand \@url [1]{\endgroup\@href {#1}{\urlprefix }}%
\providecommand \urlprefix  [0]{URL }%
\providecommand \Eprint [0]{\href }%
\providecommand \doibase [0]{http://dx.doi.org/}%
\providecommand \selectlanguage [0]{\@gobble}%
\providecommand \bibinfo  [0]{\@secondoftwo}%
\providecommand \bibfield  [0]{\@secondoftwo}%
\providecommand \translation [1]{[#1]}%
\providecommand \BibitemOpen [0]{}%
\providecommand \bibitemStop [0]{}%
\providecommand \bibitemNoStop [0]{.\EOS\space}%
\providecommand \EOS [0]{\spacefactor3000\relax}%
\providecommand \BibitemShut  [1]{\csname bibitem#1\endcsname}%
\let\auto@bib@innerbib\@empty
%</preamble>
\bibitem [{\citenamefont {Wickenbrock}\ \emph
  {et~al.}(2016{\natexlab{a}})\citenamefont {Wickenbrock}, \citenamefont
  {Zheng}, \citenamefont {Bougas}, \citenamefont {Leefer}, \citenamefont
  {Afach}, \citenamefont {Jarmola}, \citenamefont {Acosta},\ and\ \citenamefont
  {Budker}}]{Wickenbrock2016_MF}%
  \BibitemOpen
  \bibfield  {author} {\bibinfo {author} {\bibfnamefont {A.}~\bibnamefont
  {Wickenbrock}}, \bibinfo {author} {\bibfnamefont {H.}~\bibnamefont {Zheng}},
  \bibinfo {author} {\bibfnamefont {L.}~\bibnamefont {Bougas}}, \bibinfo
  {author} {\bibfnamefont {N.}~\bibnamefont {Leefer}}, \bibinfo {author}
  {\bibfnamefont {S.}~\bibnamefont {Afach}}, \bibinfo {author} {\bibfnamefont
  {A.}~\bibnamefont {Jarmola}}, \bibinfo {author} {\bibfnamefont {V.~M.}\
  \bibnamefont {Acosta}}, \ and\ \bibinfo {author} {\bibfnamefont
  {D.}~\bibnamefont {Budker}},\ }\href@noop {} {\bibfield  {journal} {\bibinfo
  {journal} {Appl. Phys. Lett.}\ }\textbf {\bibinfo {volume} {109}},\ \bibinfo
  {pages} {053505} (\bibinfo {year} {2016}{\natexlab{a}})}\BibitemShut
  {NoStop}%
\bibitem [{\citenamefont {Kucsko}\ \emph {et~al.}(2013)\citenamefont {Kucsko},
  \citenamefont {Maurer}, \citenamefont {Yao}, \citenamefont {Kubo},
  \citenamefont {Noh}, \citenamefont {Lo}, \citenamefont {Park},\ and\
  \citenamefont {Lukin}}]{Kucsko2013_tem_depen_NV}%
  \BibitemOpen
  \bibfield  {author} {\bibinfo {author} {\bibfnamefont {G.}~\bibnamefont
  {Kucsko}}, \bibinfo {author} {\bibfnamefont {P.~C.}\ \bibnamefont {Maurer}},
  \bibinfo {author} {\bibfnamefont {N.~Y.}\ \bibnamefont {Yao}}, \bibinfo
  {author} {\bibfnamefont {M.}~\bibnamefont {Kubo}}, \bibinfo {author}
  {\bibfnamefont {H.~J.}\ \bibnamefont {Noh}}, \bibinfo {author} {\bibfnamefont
  {P.~K.}\ \bibnamefont {Lo}}, \bibinfo {author} {\bibfnamefont
  {H.}~\bibnamefont {Park}}, \ and\ \bibinfo {author} {\bibfnamefont {M.~D.}\
  \bibnamefont {Lukin}},\ }\href@noop {} {\bibfield  {journal} {\bibinfo
  {journal} {Nat. Phys. Lett.}\ }\textbf {\bibinfo {volume} {500}},\ \bibinfo
  {pages} {54} (\bibinfo {year} {2013})}\BibitemShut {NoStop}%
\bibitem [{\citenamefont {Ovartchaiyapong}\ \emph {et~al.}(2014)\citenamefont
  {Ovartchaiyapong}, \citenamefont {Lee}, \citenamefont {Myers}, \citenamefont
  {Jayich}, \citenamefont {Stacey}, \citenamefont {Budker},\ and\ \citenamefont
  {Hollenberg}}]{Ovartchaiyapong2014_strain_effect_NV}%
  \BibitemOpen
  \bibfield  {author} {\bibinfo {author} {\bibfnamefont {P.}~\bibnamefont
  {Ovartchaiyapong}}, \bibinfo {author} {\bibfnamefont {K.~W.}\ \bibnamefont
  {Lee}}, \bibinfo {author} {\bibfnamefont {B.~A.}\ \bibnamefont {Myers}},
  \bibinfo {author} {\bibfnamefont {A.~C.~B.}\ \bibnamefont {Jayich}}, \bibinfo
  {author} {\bibfnamefont {A.}~\bibnamefont {Stacey}}, \bibinfo {author}
  {\bibfnamefont {D.}~\bibnamefont {Budker}}, \ and\ \bibinfo {author}
  {\bibfnamefont {L.~C.~L.}\ \bibnamefont {Hollenberg}},\ }\href@noop {}
  {\bibfield  {journal} {\bibinfo  {journal} {Nat. Commun}\ }\textbf {\bibinfo
  {volume} {5}},\ \bibinfo {pages} {4429} (\bibinfo {year} {2014})}\BibitemShut
  {NoStop}%
\bibitem [{\citenamefont {Ajoy}\ and\ \citenamefont
  {Cappellaro}(2012)}]{Ajoy2012_rotation_NV}%
  \BibitemOpen
  \bibfield  {author} {\bibinfo {author} {\bibfnamefont {A.}~\bibnamefont
  {Ajoy}}\ and\ \bibinfo {author} {\bibfnamefont {P.}~\bibnamefont
  {Cappellaro}},\ }\href@noop {} {\bibfield  {journal} {\bibinfo  {journal}
  {Phys. Rev. A.}\ }\textbf {\bibinfo {volume} {86}},\ \bibinfo {pages}
  {062104} (\bibinfo {year} {2012})}\BibitemShut {NoStop}%
\bibitem [{\citenamefont {Ledbetter}\ \emph {et~al.}(2012)\citenamefont
  {Ledbetter}, \citenamefont {Jensen}, \citenamefont {Fischer}, \citenamefont
  {Jarmola},\ and\ \citenamefont {Budker}}]{Ledbetter2012_gyroscope_NV}%
  \BibitemOpen
  \bibfield  {author} {\bibinfo {author} {\bibfnamefont {M.}~\bibnamefont
  {Ledbetter}}, \bibinfo {author} {\bibfnamefont {K.}~\bibnamefont {Jensen}},
  \bibinfo {author} {\bibfnamefont {R.}~\bibnamefont {Fischer}}, \bibinfo
  {author} {\bibfnamefont {A.}~\bibnamefont {Jarmola}}, \ and\ \bibinfo
  {author} {\bibfnamefont {D.}~\bibnamefont {Budker}},\ }\href@noop {}
  {\bibfield  {journal} {\bibinfo  {journal} {Phys. Rev. A}\ }\textbf {\bibinfo
  {volume} {86}},\ \bibinfo {pages} {052116} (\bibinfo {year}
  {2012})}\BibitemShut {NoStop}%
\bibitem [{\citenamefont {Dolde}\ \emph {et~al.}(2011)\citenamefont {Dolde},
  \citenamefont {Fedder}, \citenamefont {Doherty}, \citenamefont {Nobauer},
  \citenamefont {Rempp}, \citenamefont {Balasubramanian}, \citenamefont {Wolf},
  \citenamefont {Reinhard}, \citenamefont {Hollenberg}, \citenamefont
  {Jelezko},\ and\ \citenamefont {Wrachtrup}}]{Dolde2011_Electricfield_NV}%
  \BibitemOpen
  \bibfield  {author} {\bibinfo {author} {\bibfnamefont {F.}~\bibnamefont
  {Dolde}}, \bibinfo {author} {\bibfnamefont {H.}~\bibnamefont {Fedder}},
  \bibinfo {author} {\bibfnamefont {M.~W.}\ \bibnamefont {Doherty}}, \bibinfo
  {author} {\bibfnamefont {T.}~\bibnamefont {Nobauer}}, \bibinfo {author}
  {\bibfnamefont {F.}~\bibnamefont {Rempp}}, \bibinfo {author} {\bibfnamefont
  {G.}~\bibnamefont {Balasubramanian}}, \bibinfo {author} {\bibfnamefont
  {T.}~\bibnamefont {Wolf}}, \bibinfo {author} {\bibfnamefont {F.}~\bibnamefont
  {Reinhard}}, \bibinfo {author} {\bibfnamefont {L.~C.~L.}\ \bibnamefont
  {Hollenberg}}, \bibinfo {author} {\bibfnamefont {F.}~\bibnamefont {Jelezko}},
  \ and\ \bibinfo {author} {\bibfnamefont {J.}~\bibnamefont {Wrachtrup}},\
  }\href@noop {} {\bibfield  {journal} {\bibinfo  {journal} {nat. Phys.}\
  }\textbf {\bibinfo {volume} {7}},\ \bibinfo {pages} {459} (\bibinfo {year}
  {2011})}\BibitemShut {NoStop}%
\bibitem [{\citenamefont {Block}\ \emph {et~al.}(2020)\citenamefont {Block},
  \citenamefont {Kobrin}, \citenamefont {Jarmola}, \citenamefont {Hsieh},
  \citenamefont {Zu}, \citenamefont {Figueroa}, \citenamefont {Acosta},
  \citenamefont {Minguzzi}, \citenamefont {Maze}, \citenamefont {Budker},\ and\
  \citenamefont {Yao}}]{Block2020_Electricfield_NV}%
  \BibitemOpen
  \bibfield  {author} {\bibinfo {author} {\bibfnamefont {M.}~\bibnamefont
  {Block}}, \bibinfo {author} {\bibfnamefont {B.}~\bibnamefont {Kobrin}},
  \bibinfo {author} {\bibfnamefont {A.}~\bibnamefont {Jarmola}}, \bibinfo
  {author} {\bibfnamefont {S.}~\bibnamefont {Hsieh}}, \bibinfo {author}
  {\bibfnamefont {C.}~\bibnamefont {Zu}}, \bibinfo {author} {\bibfnamefont
  {N.~L.}\ \bibnamefont {Figueroa}}, \bibinfo {author} {\bibfnamefont {V.~M.}\
  \bibnamefont {Acosta}}, \bibinfo {author} {\bibfnamefont {J.}~\bibnamefont
  {Minguzzi}}, \bibinfo {author} {\bibfnamefont {J.~R.}\ \bibnamefont {Maze}},
  \bibinfo {author} {\bibfnamefont {D.}~\bibnamefont {Budker}}, \ and\ \bibinfo
  {author} {\bibfnamefont {N.~Y.}\ \bibnamefont {Yao}},\ }\href@noop {}
  {\bibfield  {journal} {\bibinfo  {journal} {arXiv}\ }\textbf {\bibinfo
  {volume} {1}},\ \bibinfo {pages} {02886} (\bibinfo {year}
  {2020})}\BibitemShut {NoStop}%
\bibitem [{\citenamefont {Wickenbrock}\ \emph {et~al.}(2014)\citenamefont
  {Wickenbrock}, \citenamefont {Jurgilas}, \citenamefont {Dow}, \citenamefont
  {Marmugi},\ and\ \citenamefont {Renzoni}}]{Wickenbrock2014_vaporcell}%
  \BibitemOpen
  \bibfield  {author} {\bibinfo {author} {\bibfnamefont {A.}~\bibnamefont
  {Wickenbrock}}, \bibinfo {author} {\bibfnamefont {S.}~\bibnamefont
  {Jurgilas}}, \bibinfo {author} {\bibfnamefont {A.}~\bibnamefont {Dow}},
  \bibinfo {author} {\bibfnamefont {L.}~\bibnamefont {Marmugi}}, \ and\
  \bibinfo {author} {\bibfnamefont {R.}~\bibnamefont {Renzoni}},\ }\href@noop
  {} {\bibfield  {journal} {\bibinfo  {journal} {Optics. Lett.}\ }\textbf
  {\bibinfo {volume} {39}},\ \bibinfo {pages} {6367} (\bibinfo {year}
  {2014})}\BibitemShut {NoStop}%
\bibitem [{\citenamefont {Marmugi}\ \emph {et~al.}(2016)\citenamefont
  {Marmugi}, \citenamefont {Hussain},\ and\ \citenamefont
  {Renzoni}}]{Marmugi2016_vaporcell}%
  \BibitemOpen
  \bibfield  {author} {\bibinfo {author} {\bibfnamefont {L.}~\bibnamefont
  {Marmugi}}, \bibinfo {author} {\bibfnamefont {S.}~\bibnamefont {Hussain}}, \
  and\ \bibinfo {author} {\bibfnamefont {F.}~\bibnamefont {Renzoni}},\
  }\href@noop {} {\bibfield  {journal} {\bibinfo  {journal} {Appl. Phys.
  Lett.}\ }\textbf {\bibinfo {volume} {108}},\ \bibinfo {pages} {103503}
  (\bibinfo {year} {2016})}\BibitemShut {NoStop}%
\bibitem [{\citenamefont {Wickenbrock}\ \emph
  {et~al.}(2016{\natexlab{b}})\citenamefont {Wickenbrock}, \citenamefont
  {Leefer}, \citenamefont {Blanchard},\ and\ \citenamefont
  {Budker}}]{Wickenbrock2016_vaporcelleddy}%
  \BibitemOpen
  \bibfield  {author} {\bibinfo {author} {\bibfnamefont {A.}~\bibnamefont
  {Wickenbrock}}, \bibinfo {author} {\bibfnamefont {N.}~\bibnamefont {Leefer}},
  \bibinfo {author} {\bibfnamefont {J.~W.}\ \bibnamefont {Blanchard}}, \ and\
  \bibinfo {author} {\bibfnamefont {D.}~\bibnamefont {Budker}},\ }\href@noop {}
  {\bibfield  {journal} {\bibinfo  {journal} {Appl. Phys. Lett.}\ }\textbf
  {\bibinfo {volume} {108}},\ \bibinfo {pages} {183507} (\bibinfo {year}
  {2016}{\natexlab{b}})}\BibitemShut {NoStop}%
\bibitem [{\citenamefont {Chatzidrosos}\ \emph {et~al.}(2019)\citenamefont
  {Chatzidrosos}, \citenamefont {Wickenbrock}, \citenamefont {Bougas},
  \citenamefont {Zheng}, \citenamefont {Tretiak}, \citenamefont {Yang},\ and\
  \citenamefont {Budker}}]{Chatzifrosos2019_eddycurrent_NV}%
  \BibitemOpen
  \bibfield  {author} {\bibinfo {author} {\bibfnamefont {G.}~\bibnamefont
  {Chatzidrosos}}, \bibinfo {author} {\bibfnamefont {A.}~\bibnamefont
  {Wickenbrock}}, \bibinfo {author} {\bibfnamefont {L.}~\bibnamefont {Bougas}},
  \bibinfo {author} {\bibfnamefont {H.}~\bibnamefont {Zheng}}, \bibinfo
  {author} {\bibfnamefont {O.}~\bibnamefont {Tretiak}}, \bibinfo {author}
  {\bibfnamefont {Y.}~\bibnamefont {Yang}}, \ and\ \bibinfo {author}
  {\bibfnamefont {D.}~\bibnamefont {Budker}},\ }\href@noop {} {\bibfield
  {journal} {\bibinfo  {journal} {Phys. Rev. Appl.}\ }\textbf {\bibinfo
  {volume} {11}},\ \bibinfo {pages} {014060} (\bibinfo {year}
  {2019})}\BibitemShut {NoStop}%
\bibitem [{\citenamefont {Acosta}\ \emph
  {et~al.}(2010{\natexlab{a}})\citenamefont {Acosta}, \citenamefont {Bauch},
  \citenamefont {Jarmola}, \citenamefont {Zipp}, \citenamefont {Ledbetter},\
  and\ \citenamefont {Budker}}]{Acosta2010_ISC}%
  \BibitemOpen
  \bibfield  {author} {\bibinfo {author} {\bibfnamefont {V.~M.}\ \bibnamefont
  {Acosta}}, \bibinfo {author} {\bibfnamefont {E.}~\bibnamefont {Bauch}},
  \bibinfo {author} {\bibfnamefont {A.}~\bibnamefont {Jarmola}}, \bibinfo
  {author} {\bibfnamefont {L.~J.}\ \bibnamefont {Zipp}}, \bibinfo {author}
  {\bibfnamefont {M.~P.}\ \bibnamefont {Ledbetter}}, \ and\ \bibinfo {author}
  {\bibfnamefont {D.}~\bibnamefont {Budker}},\ }\href@noop {} {\bibfield
  {journal} {\bibinfo  {journal} {Appl. Phys. Lett.}\ }\textbf {\bibinfo
  {volume} {97}},\ \bibinfo {pages} {174104} (\bibinfo {year}
  {2010}{\natexlab{a}})}\BibitemShut {NoStop}%
\bibitem [{\citenamefont {Zheng}\ \emph {et~al.}(2017)\citenamefont {Zheng},
  \citenamefont {Chatzidrosos}, \citenamefont {Wickenbrock}, \citenamefont
  {Bougas}, \citenamefont {Lazda}, \citenamefont {Berzins}, \citenamefont
  {Gahbauer}, \citenamefont {Auzinsh}, \citenamefont {Ferber},\ and\
  \citenamefont {Budker}}]{Zheng2017_gslac}%
  \BibitemOpen
  \bibfield  {author} {\bibinfo {author} {\bibfnamefont {H.}~\bibnamefont
  {Zheng}}, \bibinfo {author} {\bibfnamefont {G.}~\bibnamefont {Chatzidrosos}},
  \bibinfo {author} {\bibfnamefont {A.}~\bibnamefont {Wickenbrock}}, \bibinfo
  {author} {\bibfnamefont {L.}~\bibnamefont {Bougas}}, \bibinfo {author}
  {\bibfnamefont {R.}~\bibnamefont {Lazda}}, \bibinfo {author} {\bibfnamefont
  {A.}~\bibnamefont {Berzins}}, \bibinfo {author} {\bibfnamefont {F.~H.}\
  \bibnamefont {Gahbauer}}, \bibinfo {author} {\bibfnamefont {M.}~\bibnamefont
  {Auzinsh}}, \bibinfo {author} {\bibfnamefont {R.}~\bibnamefont {Ferber}}, \
  and\ \bibinfo {author} {\bibfnamefont {D.}~\bibnamefont {Budker}},\
  }\href@noop {} {\bibfield  {journal} {\bibinfo  {journal} {Proc. SPIE}\
  }\textbf {\bibinfo {volume} {10119}},\ \bibinfo {pages} {101190X} (\bibinfo
  {year} {2017})}\BibitemShut {NoStop}%
\bibitem [{\citenamefont {Hall}\ \emph {et~al.}(2016)\citenamefont {Hall},
  \citenamefont {Kehayias}, \citenamefont {Simpson}, \citenamefont {Jarmola},
  \citenamefont {Stacey}, \citenamefont {Budker},\ and\ \citenamefont
  {Hollenberg}}]{Hall2016_features_NV}%
  \BibitemOpen
  \bibfield  {author} {\bibinfo {author} {\bibfnamefont {L.~T.}\ \bibnamefont
  {Hall}}, \bibinfo {author} {\bibfnamefont {P.}~\bibnamefont {Kehayias}},
  \bibinfo {author} {\bibfnamefont {D.~A.}\ \bibnamefont {Simpson}}, \bibinfo
  {author} {\bibfnamefont {A.}~\bibnamefont {Jarmola}}, \bibinfo {author}
  {\bibfnamefont {A.}~\bibnamefont {Stacey}}, \bibinfo {author} {\bibfnamefont
  {D.}~\bibnamefont {Budker}}, \ and\ \bibinfo {author} {\bibfnamefont
  {L.~C.~L.}\ \bibnamefont {Hollenberg}},\ }\href@noop {} {\bibfield  {journal}
  {\bibinfo  {journal} {Nat. Commun.}\ }\textbf {\bibinfo {volume} {7}},\
  \bibinfo {pages} {10211} (\bibinfo {year} {2016})}\BibitemShut {NoStop}%
\bibitem [{\citenamefont {Anishchik}\ and\ \citenamefont
  {Ivanov}(2017)}]{Anishchik2017_features_NV}%
  \BibitemOpen
  \bibfield  {author} {\bibinfo {author} {\bibfnamefont {S.~V.}\ \bibnamefont
  {Anishchik}}\ and\ \bibinfo {author} {\bibfnamefont {K.~L.}\ \bibnamefont
  {Ivanov}},\ }\href@noop {} {\bibfield  {journal} {\bibinfo  {journal} {Phys.
  Rev. B}\ }\textbf {\bibinfo {volume} {96}},\ \bibinfo {pages} {115142}
  (\bibinfo {year} {2017})}\BibitemShut {NoStop}%
\bibitem [{\citenamefont {Acosta}\ \emph
  {et~al.}(2010{\natexlab{b}})\citenamefont {Acosta}, \citenamefont {Bauch},
  \citenamefont {Ledbetter}, \citenamefont {Waxman}, \citenamefont {Bouchard},\
  and\ \citenamefont {Budker}}]{Acosta2010_temdepen_NV}%
  \BibitemOpen
  \bibfield  {author} {\bibinfo {author} {\bibfnamefont {V.~M.}\ \bibnamefont
  {Acosta}}, \bibinfo {author} {\bibfnamefont {E.}~\bibnamefont {Bauch}},
  \bibinfo {author} {\bibfnamefont {M.~P.}\ \bibnamefont {Ledbetter}}, \bibinfo
  {author} {\bibfnamefont {A.}~\bibnamefont {Waxman}}, \bibinfo {author}
  {\bibfnamefont {L.-S.}\ \bibnamefont {Bouchard}}, \ and\ \bibinfo {author}
  {\bibfnamefont {D.}~\bibnamefont {Budker}},\ }\href@noop {} {\bibfield
  {journal} {\bibinfo  {journal} {Phys. Rev. Lett.}\ }\textbf {\bibinfo
  {volume} {104}},\ \bibinfo {pages} {070801} (\bibinfo {year}
  {2010}{\natexlab{b}})}\BibitemShut {NoStop}%
\bibitem [{\citenamefont {AG}()}]{TDKweb}%
  \BibitemOpen
  \bibfield  {author} {\bibinfo {author} {\bibfnamefont {T.~E.}\ \bibnamefont
  {AG}},\ }\href
  {https://www.tdk-electronics.tdk.com/download/2427688/34992408a5ac6711037617262201d5b0/06-dl-ceracharge-presentation.pdf}
  {\enquote {\bibinfo {title} {Ceracharge - rechargeable solid-state smd
  battery},}\ }\BibitemShut {NoStop}%
\bibitem [{\citenamefont {Hu}\ \emph {et~al.}(2020{\natexlab{a}})\citenamefont
  {Hu}, \citenamefont {Iwata}, \citenamefont {Bougas}, \citenamefont
  {Blanchard}, \citenamefont {Wickenbrock}, \citenamefont {Jakob},
  \citenamefont {Schwarz}, \citenamefont {Schwarzinger}, \citenamefont
  {Jerschow},\ and\ \citenamefont {Budker}}]{hu2020rapid}%
  \BibitemOpen
  \bibfield  {author} {\bibinfo {author} {\bibfnamefont {Y.}~\bibnamefont
  {Hu}}, \bibinfo {author} {\bibfnamefont {G.~Z.}\ \bibnamefont {Iwata}},
  \bibinfo {author} {\bibfnamefont {L.}~\bibnamefont {Bougas}}, \bibinfo
  {author} {\bibfnamefont {J.~W.}\ \bibnamefont {Blanchard}}, \bibinfo {author}
  {\bibfnamefont {A.}~\bibnamefont {Wickenbrock}}, \bibinfo {author}
  {\bibfnamefont {G.}~\bibnamefont {Jakob}}, \bibinfo {author} {\bibfnamefont
  {S.}~\bibnamefont {Schwarz}}, \bibinfo {author} {\bibfnamefont
  {C.}~\bibnamefont {Schwarzinger}}, \bibinfo {author} {\bibfnamefont
  {A.}~\bibnamefont {Jerschow}}, \ and\ \bibinfo {author} {\bibfnamefont
  {D.}~\bibnamefont {Budker}},\ }\href@noop {} {\bibfield  {journal} {\bibinfo
  {journal} {Applied Sciences}\ }\textbf {\bibinfo {volume} {10}},\ \bibinfo
  {pages} {7864} (\bibinfo {year} {2020}{\natexlab{a}})}\BibitemShut {NoStop}%
\bibitem [{\citenamefont {Gomory}(1997)}]{Gomory1997_ACsusceptibility}%
  \BibitemOpen
  \bibfield  {author} {\bibinfo {author} {\bibfnamefont {F.}~\bibnamefont
  {Gomory}},\ }\href@noop {} {\bibfield  {journal} {\bibinfo  {journal}
  {Supercond. Sci. Technol.}\ }\textbf {\bibinfo {volume} {10}},\ \bibinfo
  {pages} {523} (\bibinfo {year} {1997})}\BibitemShut {NoStop}%
\bibitem [{\citenamefont {Mulder}\ \emph {et~al.}(1981)\citenamefont {Mulder},
  \citenamefont {van Duyneveldt},\ and\ \citenamefont
  {Mydosh}}]{Mulder1981_ACsusceptibility}%
  \BibitemOpen
  \bibfield  {author} {\bibinfo {author} {\bibfnamefont {C.~A.~M.}\
  \bibnamefont {Mulder}}, \bibinfo {author} {\bibfnamefont {A.~J.}\
  \bibnamefont {van Duyneveldt}}, \ and\ \bibinfo {author} {\bibfnamefont
  {J.~A.}\ \bibnamefont {Mydosh}},\ }\href@noop {} {\bibfield  {journal}
  {\bibinfo  {journal} {Phys. Rev. B}\ }\textbf {\bibinfo {volume} {23}},\
  \bibinfo {pages} {1384} (\bibinfo {year} {1981})}\BibitemShut {NoStop}%
\bibitem [{\citenamefont {Hu}\ \emph {et~al.}(2020{\natexlab{b}})\citenamefont
  {Hu}, \citenamefont {Iwata}, \citenamefont {Mohammadi}, \citenamefont
  {Silletta}, \citenamefont {Wickenbrock}, \citenamefont {Blanchard},
  \citenamefont {Budker},\ and\ \citenamefont {Jerschow}}]{hu2020sensitive}%
  \BibitemOpen
  \bibfield  {author} {\bibinfo {author} {\bibfnamefont {Y.}~\bibnamefont
  {Hu}}, \bibinfo {author} {\bibfnamefont {G.~Z.}\ \bibnamefont {Iwata}},
  \bibinfo {author} {\bibfnamefont {M.}~\bibnamefont {Mohammadi}}, \bibinfo
  {author} {\bibfnamefont {E.~V.}\ \bibnamefont {Silletta}}, \bibinfo {author}
  {\bibfnamefont {A.}~\bibnamefont {Wickenbrock}}, \bibinfo {author}
  {\bibfnamefont {J.~W.}\ \bibnamefont {Blanchard}}, \bibinfo {author}
  {\bibfnamefont {D.}~\bibnamefont {Budker}}, \ and\ \bibinfo {author}
  {\bibfnamefont {A.}~\bibnamefont {Jerschow}},\ }\href@noop {} {\bibfield
  {journal} {\bibinfo  {journal} {Proceedings of the National Academy of
  Sciences}\ }\textbf {\bibinfo {volume} {117}},\ \bibinfo {pages} {10667}
  (\bibinfo {year} {2020}{\natexlab{b}})}\BibitemShut {NoStop}%
\end{thebibliography}%

\end{document}